\documentstyle[twocolumn,prl,aps,epsf]{revtex} 
\newcommand{\be}{\begin{eqnarray}}
\newcommand{\ee}{\end{eqnarray}}
\def\Dirac#1{#1\hskip-5pt/}

\begin{document}
\draft
\title{Hard exclusive electroproduction of decuplet baryons 
in the large $N_c$ limit}

\author{L.L. Frankfurt$^1$, M.V. Polyakov$^{2,3}$, 
M. Strikman$^4$ and M. Vanderhaeghen$^5$}
\address{$^1$ School of Physics and Astronomy, Tel Aviv University, 69978 
Tel Aviv, Israel}
\address{$^2$ Petersburg Nuclear Physics Institute, 188350 Gatchina, Russia}
\address{$^3$ Institut f\"ur Theoretische Physik II, Ruhr-Universit\"at
  Bochum, D-44780 Bochum, Germany}
\address{$^4$ Department of Physics, Pennsylvania State University,
University Park, PA 16802, USA}
\address{$^5$ Institut f\"ur Kernphysik, Johannes Gutenberg Universit\"at,
  D-55099 Mainz, Germany}
\date{\today}
\maketitle

\begin{abstract}
The cross sections and transverse spin asymmetries in the  
hard exclusive electroproduction of decuplet baryons 
are calculated in the large $N_c$ limit 
and found to be comparable to that of octet baryons. 
Large $N_c$ selection rules for the production amplitudes are derived, 
leading to  new sensitive tests
 of the spin aspects of the QCD chiral dynamics both in the nonstrange
 and strange sectors.
Importance of such studies for the reliable extraction
of the pion form factor from pion electroproduction is explained.
\end{abstract}
\begin{figure}[h]
\epsfxsize=7.5 cm
\epsfysize=9. cm
\centerline{\epsffile{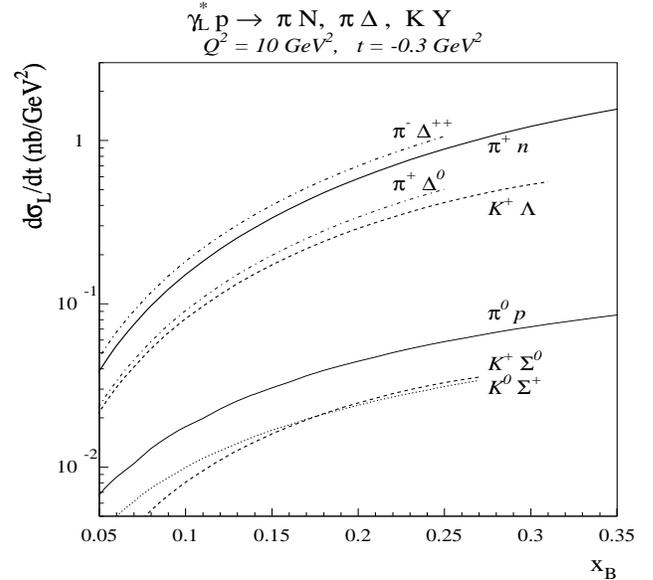}}
\vspace{-.8cm}
\caption[]{\small Leading order predictions for the 
$\pi N$, $\pi \Delta$ and $K Y$ 
longitudinal electroproduction cross sections 
at $t$ = -0.3~GeV$^2$, as function of $x_B$
(plotted up to the $x_B$ value for which $t = t_{min}$). 
Results for pion and charged kaon channels are given using an
asymptotic DA. For $K^0 \Sigma^+$, 
predictions are shown using the CZ DA with antisymmetric
part : $\eta^a_K = 0.25$.}
\label{fig:pionkhyp}
\end{figure}

\narrowtext
Recently the QCD factorization theorem was shown to hold \cite{CFS}
 for a wide range of two-body
exclusive deep inelastic processes
with $t = \Delta^2$ and $x_B=Q^{2}/2p\cdot q$ fixed:
\begin{equation}
     \gamma_L^{*}(q) + p \to  M(q - \Delta ) + B'(p + \Delta ).
\label{process}
\end{equation}
It asserts that at sufficiently large $Q^2$, the amplitude has the form~:
\begin{eqnarray}
   &&
   \sum _{i,j} \int _{0}^{1}dz  \int d x_1 \,
   f_{i/p}(x_1 ,x_1 -x_B;t,\mu ) \, \nonumber\\
&& \hspace{2.cm} \times \, H_{ij}(Q^{2}x_1 /x_B,Q^{2},z,\mu )
   \, \phi _{j}(z,\mu )
\nonumber\\
&&
   +\, \mbox{power-suppressed corrections} .
\label{factorization}
\end{eqnarray}
The nonzero momentum transfer $\Delta$ for forward angle scattering 
introduces a skewedness $\xi$, which is defined through the light cone fraction
$\Delta^+ = -(2 \xi) P^+$
(with $P=(p_p+p_{B'})/2$ - the average four-momentum of the
initial nucleon and the recoiling hadron B').
In Eq.~(\ref{factorization}), $f$ are the skewed parton distributions
(SPD) (see formal definition below),
 $\phi $ is the minimal Fock component of the
 light-front wave function of the
meson, and $H$ is a hard-scattering coefficient, 
computable in powers of $\alpha _{s}(Q)$. By virtue of the
factorization theorem, SPD's are universal and enter in a wide range of hard
exclusive two body processes. Model calculations of SPD's are
currently possible within the QCD chiral models for intermediate $x_B$
\cite{PPP97,PPG}.
One may expect that eventually it would be possible to calculate SPD's
for intermediate $x_B$ using lattice QCD.
\newline
\indent
Hence reactions (\ref{process})
 provide a new tool to probe the
quark-gluon structure of various mesons and baryons as well as the
hard QCD dynamics. Many studies focused on the production of vector
mesons at HERA energies where this process is primarily sensitive
to the gluon dynamics in nucleons at small $x_B$ and to the properties of
the vector meson
wave functions. Studies at larger $x_{B}$ - at 
lower energies - provide a wide range of new opportunities.
One of the most promising ones theoretically and feasible
experimentally is the production of pseudoscalar
mesons. So far the main focus was 
the reaction $\gamma_L^* + N \to \pi(\eta) + N'$
\cite{MPR,FPPS,EFS} 
where $N'$ is an octet baryon \cite{FPPS}.
The analysis  based on the  chiral QCD  dynamics has demonstrated that
the SPD's which disappear in the
case of diagonal parton densities play an important role in these
processes, leading in particular to a strong dependence of the
differential cross section on the transverse polarization of the target.
\newline
\indent
One of the intriguing questions of medium-energy QCD dynamics is
the differences and similarities in the structure of  baryons belonging 
to the different $SU(3)_f $ multiplets. 
In particular, a naive constituent quark model suggests that they are
similar, while there are suggestions that due to a strong attraction
between the quarks in the spin-isospin zero channel, diquark 
correlations should be important in the octet baryons but not in the
decuplet \cite{Suryak}.
At the same time
the chiral models suggest that in the large
$N_c$ limit, nucleons and
$\Delta$ isobars are different
rotational excitations of the same soliton \cite{soliton}.
\newline
\indent
Clearly an ability to compare experimentally
the wave functions of the decuplet and
octet would be very helpul to discriminate between different ideas.
So far only one process could be used for these purposes -
electroexcitation of $\Delta$ isobars \cite{Stoler}.
Here we want to explore the
potential of the process $\gamma^*_L +N \to \pi +\Delta $
as well as the DVCS process $\gamma^* +N \to \gamma +\Delta $ for these
studies. In the experiments with low resolution 
in the mass of the recoiling system ($\Delta M \approx$ 300 MeV for
HERMES in the current set-up),  
the estimates of $\Delta$ production are necessary
to extract the $N \to N$ SPD's from such data.
Note also that the study of these processes would allow to make a more
reliable separation of the pion pole contribution in the
electroproduction of pions which is mandatory for the measurement of
the pion elastic form factor.
We use the large $N_c$, 
to predict a number of
striking characteristics of these processes. In particular we predict
large absolute cross sections for a number of channels
including those with production of octet strange baryons 
(extension to decuplet strange baryons will be considered in a future work). 
We also predict large transverse spin asymmetries 
for $\Delta $ and $\Lambda$ production, 
related to the peculiar feature of the chiral QCD.
Hence study of these processes can provide unique tests of the
soliton type  approach to baryon structure. 
We will mostly focus on the ratios of cross sections for the 
different channels and on the spin asymmetries which are likely to be less
sensitive to higher twist effects and hence could be explored
already using the HERMES detector and TJNAF at higher energies.
We hope that this letter will encourage upgrades of the
detector recoil capabilities which are necessary
for the study of some of the discussed effects as well as 
measurements with transversely polarized targets.
\newline
\indent
We define a new set of SPD's for the axial $N\to \Delta$ (isovector) 
transition, denoted as the skewed distributions
$C_i^{(3)}$ (functions of $x, \xi$ and $\Delta^2$), which enter into 
$\pi \Delta$ electroproduction~:
\begin{eqnarray}
&&\int \frac{d\lambda }{2\pi }e^{i\lambda x}\langle
\Delta^+, p^{\prime }|\bar \psi
(-\lambda n/2){\Dirac n}\gamma^5
 \psi (\lambda n/2)|N, p\rangle = \nonumber \\
&& \bar \psi^\beta(p') \; \bigl[\,
C_1^{(3)} n_\beta + C_2^{(3)} \frac{\Delta_\beta(n\cdot\Delta)}{m_N^2}
+\ldots \bigr] \; u(p) ,
\label{eq:axialndelta}
\end{eqnarray}
where $\psi^\beta(p')$ is the Rarita-Schwinger spinor and ellipses
denote other contributions which are suppressed at large $N_c$ 
(details will be published elsewhere \cite{PV}). 
In the large $N_c$ limit, the nucleon
and $\Delta$ are rotational excitations of {\em the same} classical
object-soliton. This allows to derive a number of relations between
$N\to N$ and $N\to\Delta$ SPD's. For $C_1^{(3)}$ and $C_2^{(3)}$, 
they have the form:
\begin{eqnarray}
\label{eq:spdndelta1}
C_1^{(3)}(x, \xi, \Delta^2) \,&=&\, \sqrt{3} \,
\widetilde H^{(3)}(x, \xi, \Delta^2) \, ,\\
C_2^{(3)}(x, \xi, \Delta^2) \,&=&\, \sqrt{3}/4 \,
\widetilde E^{(3)}(x, \xi, \Delta^2) \, ,
\label{eq:spdndelta2}
\end{eqnarray}
where the SPD $\widetilde H^{(3)} = \widetilde H^u - \widetilde
H^d$ and analogously for $\widetilde E^{(3)}$.           
We use notations of Ji \cite{JiReview} for the $N\to N$ SPD's. 
\newline
\indent
For the $N \to \Delta$ DVCS process, besides the {\it axial} 
SPD's of Eq.~(\ref{eq:axialndelta}), also {\it vector} 
SPD's enter, which are defined as~:
\be
\nonumber
&&\int \frac{d\lambda }{2\pi }e^{i\lambda x}\langle
\Delta^+, p^{\prime }|\bar \psi
(-\lambda n/2){\Dirac n}
 \psi (\lambda n/2)|N, p\rangle =\\
&&\sqrt{\frac{2}{3}} \; \bar \psi^\beta(p') \; \bigl[
H_M^{(3)}(x,\xi ,\Delta^2) \, {\cal K}_{\beta\mu}^M n^\mu + ... \bigr]
\; u(p) \, , 
\label{eq:spdndelta3}
\ee
where we show only the dominant (at large $N_c$) magnetic 
contribution $H_M^{(3)}$ (${\cal K}_{\beta\mu}^M$ is the 
corresponding Lorentz structure, see \cite{PV} for details). 
At large $N_c$, we find~:
\begin{equation}
H_M^{(3)}(x, \xi, \Delta^2) = 2/\sqrt{3} \,
E^{(3)}(x, \xi, \Delta^2) \, ,
\label{eq:spdndelta4}
\end{equation}
in terms of the isovector unpolarized $N \to N$ SPD $E^{(3)}$, 
giving comparable $\gamma^* N\to\gamma N$ and $\gamma^* N\to\gamma
\Delta$ amplitudes 
(except that the SPD $H$ \cite{JiReview} is absent for $N\to \Delta$).
\newline
\indent
Using the large $N_c$ relations of 
Eqs.~(\ref{eq:spdndelta1},\ref{eq:spdndelta2}), one can easily
derive the relations between the different cross sections for charged
pion production as 
$\sigma_L^{\gamma^* p \to \pi^+ n}:\sigma_L^{\gamma^* p \to \pi^+\Delta^0}:\sigma_L^{\gamma^* p \to \pi^-\Delta^{++}}:\sigma_L^{\gamma^* n \to \pi^- p}
\approx 1:0.5:1.25:0.8$, see also Fig.~\ref{fig:pionkhyp}.
For the production of neutral pseudoscalar mesons we estimated
$\sigma^{\gamma^* p \to \eta(\eta')\Delta^+}_L  \ll
 \sigma_L^{\gamma^* p \to \eta(\eta') p}$ , 
and $\sigma^{\gamma^* p \to \pi^0\Delta^+}_L  \approx      
0.1 \; \sigma_L^{\gamma^* p \to \pi^0 p}$. 

\vspace{-0.5cm}
\begin{figure}[h]
\epsfxsize=7.5 cm
\epsfysize=9. cm
\centerline{\epsffile{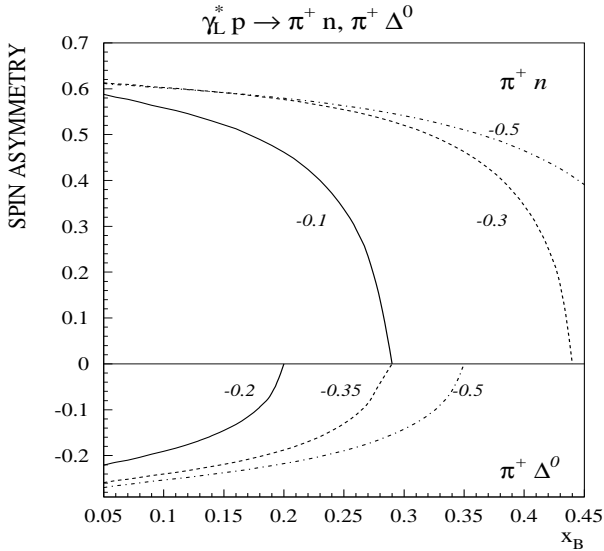}}
\vspace{-1.15cm}
\caption[]{\small Transverse spin asymmetry 
  for the longitudinal electroproduction of 
$\pi^+ n$ and $\pi^+ \Delta^0$, 
at different values of $t$ (indicated on the curves in (GeV/c)$^2$).}
\label{fig:pidelasymm}
\end{figure}

\indent
Besides the cross section $\sigma_L$, the second
observable which involves only longitudinal amplitudes and which is
a leading order observable for hard exclusive
meson electroproduction, is the single spin asymmetry, ${\cal A}_{\pi B}$ 
for a proton target polarized perpendicular to the reaction
plane (or the equivalent recoil polarization observable). 
For the hard electroproduction of $\pi N$ final states, a 
large value of ${\cal A}_{\pi N}$ was predicted 
in \cite{FPPS}. 
We give here for the first time ${\cal A}_{\pi \Delta}$ and ${\cal A}_{KY}$, 
which also turn out to be large.
\newline
\indent 
For $\pi \Delta$ final states we find~:
\begin{eqnarray}                                         
\label{eq:pidelasymm}
&&{\cal A}_{\pi \Delta} = - \, {{2 \, |\Delta_\perp|} \over {\pi}} \,
\frac{Im(A B^*) \, 2 \xi m_N^2 m_\Delta}{D_{\pi \Delta}} \, , \\
&&D_{\pi \Delta} = |A|^2 \, m_N^4 (1-\xi)^2 \nonumber\\
&&+ |B|^2 \, \xi^2 \left[ t^2 - 2 t (m_\Delta^2 + m_N^2)
+ (m_\Delta^2 - m_N^2)^2\right]  \nonumber\\
&&+ Re(AB^*) 2 \xi m_N^2 \left[ \xi t - \xi (3 m_\Delta^2 + m_N^2) -
t - m_\Delta^2 + m_N^2 \right], \nonumber
\end{eqnarray}
For $\pi^+ \Delta^0$ electroproduction , $A$ and $B$ are given by~:
\begin{eqnarray}
\label{eq:apipdelo}
A_{\pi^+ \Delta^0} \,&=&\, \int_{-1}^1 dx \, C_1^{(3)}
\left\{ {{e_u} \over {x - \xi + i \epsilon}}
+ {{e_d} \over {x + \xi - i \epsilon}} \right\} , \\
B_{\pi^+ \Delta^0} \,&=&\, \int_{-1}^1 dx \, C_2^{(3)}
\left\{ {{e_u} \over {x - \xi + i \epsilon}}
+ {{e_d} \over {x + \xi - i \epsilon}} \right\} ,
\label{eq:bpipdelo}
\end{eqnarray}
where the functions $C_1^{(3)}$, $C_2^{(3)}$ are as defined in 
Eqs.~(\ref{eq:spdndelta1},\ref{eq:spdndelta2}).
Eq.~(\ref{eq:spdndelta2}) implies that the pion pole contribution to
$B_{\pi^+ \Delta^0}$ is given by
$B^{pole}_{\pi^+ \Delta^0} = \sqrt{3}/4 \, B^{pole}_{\pi^+ n}$. 
We use an asymptotic pion distribution amplitude (DA), for which  
$B^{pole}_{\pi^+ n}$ is given by~:
\begin{equation}
B^{pole}_{\pi^+ n} = - 3 / (2 \xi) 
\,g_A \, (2 m_N)^2 / (-t + m_\pi^2)\, ,
\end{equation}
with $g_A \simeq 1.267$.
For electroproduction of $\pi^- \Delta^{++}$ one has 
- up to a global isospin factor - analogous expressions as
Eqs.~(\ref{eq:apipdelo},\ref{eq:bpipdelo}), by making the replacement
$e_u \leftrightarrow e_d$.
\newline
\indent
In Fig.~\ref{fig:pidelasymm}, we plot ${\cal A}_{\pi^+n}$ and
${\cal A}_{\pi^+\Delta^0}$ at several 
$t$, as a function of $x_B$.
We use the large $N_c$ relations 
Eqs.~(\ref{eq:spdndelta1},\ref{eq:spdndelta2}) for the 
$N \to \Delta$ SPD's. For the $N \to N$ SPD's, we use the
phenomenological $\xi$-dependent ansatz used in \cite{VGG99}. 
Fig.~\ref{fig:pidelasymm} shows that 
the predicted ${\cal A}_{\pi^+n}$ is large and comparable with
the results of \cite{FPPS} in which the SPD's computed in the chiral
quark soliton model \cite{PPG} were used. 
Fig.~\ref{fig:pidelasymm} furthermore shows that ${\cal A}_{\pi^+\Delta^0}$ 
has the opposite 
sign compared with ${\cal A}_{\pi^+n}$.
Although the magnitude of ${\cal A}_{\pi^+\Delta^0}$ 
is smaller than ${\cal A}_{\pi^+n}$, 
as anticipated in \cite{FPPS}, it is still sizeable. 
Similarly, we find 
${\cal A}_{\pi^-\Delta^{++}}\approx 0.5 \, {\cal A}_{\pi^+\Delta^0}$.
\vspace{-0.65cm}
\indent
Measurement of ${\cal A}_{\pi N}$
can provide an important help in the
extraction of the pion form factor. 
For $x_B \leq 0.15$ where one reaches values 
$t_{min} \sim 2 m_\pi^2$, we find that the  
pion pole constitutes about 70 \% to the longitudinal cross section.
Measurement of the asymmetry, which is an interference between 
the pseudoscalar (PS) and pseudovector (PV) 
contributions, would help to constrain the non-pole term, and 
in this way help to get a more reliable extraction of the pion form factor. 
\newline
\indent
We next turn to strangeness hard electroproduction reactions. 
In leading order, the charged kaon electroproduction 
channels involve the amplitudes :
\begin{eqnarray}
\label{eq:akpy}
A_{K^+ Y} &=& \int_{-1}^1 dx \widetilde H^{p \rightarrow Y}
\left\{ {{e_u} \over {x - \xi + i \epsilon}}
+ {{e_s} \over {x + \xi - i \epsilon}} \right\} , \\
B_{K^+ Y} &=& \int_{-1}^1 dx \widetilde E^{p \rightarrow Y}
\left\{ {{e_u} \over {x - \xi + i \epsilon}}
+ {{e_s} \over {x + \xi - i \epsilon}} \right\} ,
\label{eq:bkpy}
\end{eqnarray}
where $e_s$ is the $s$-quark charge.
To estimate the SPD $\widetilde H^{p \to Y}$, 
we use the $SU(3)$ relations of \cite{FPPS}.
The SPD $\widetilde E^{p \to Y}$  
contains a charged kaon pole contribution, given by~:
\begin{equation}
B^{pole}_{K^+ Y} = - 3 / (2 \xi) \, \eta^s_K
\, f_K \, g_{KNY}\, (2 m_N) / (-t + m_K^2) \, ,
\label{eq:kpole}
\end{equation}
with $f_K \simeq$ 159 MeV, and where $g_{KNY}$ are the $KNY$ coupling
constants. In line with our use of $SU(3)$ relations to estimate
$\tilde H$, we use also $SU(3)$ predictions for the coupling
constants~: $g_{K N \Lambda}/\sqrt{4 \pi} \approx -3.75$
and $g_{K N \Sigma}/\sqrt{4 \pi} \approx 1.09$, which 
are compatible with those obtained from a Regge fit to high energy kaon
photoproduction data \cite{GLV}. 
In Eq.~(\ref{eq:kpole}), $\eta^s_K$ is defined as~:
\begin{equation}
\left\{ \begin{array}{c} \eta^s_K \\ 
\eta^a_K  \\ \end{array} \right\} 
\,\equiv\, {2 \over 3} \, \int _{-1}^{+1}d \zeta \,
\left\{ \begin{array}{c} \Phi^s_K\left(\zeta\right) \\ 
\zeta \; \Phi^a_K\left(\zeta\right) \\ \end{array} \right\} 
\, {1 \over {1 - \zeta^2}} \, ,
\label{eq:czkaon}
\end{equation}
where $\Phi^s_K(\zeta)$ is the symmetric (in $\zeta$) part of the kaon
DA. In Eq.~(\ref{eq:czkaon}), we have also defined - for further use -
$\eta^a_K $ for the antisymmetric part $\Phi^a_K(\zeta)$ of the kaon DA. 
The latter is due to $SU(3)_f$ symmetry breaking effects. 
One has $\eta^s_K$ = 1 for an asymptotic DA
[$\Phi^s(\zeta) = 3/4 (1 - \zeta^2)$] and
$\eta^s_K$ = 7/5 for the CZ kaon DA \cite{CZ}.
\newline
\indent
For the $K^0 \Sigma^+$ electroproduction, we give the expressions for
$A$ and $B$, allowing for both a symmetric and antisymmetric component
in the kaon DA, which yields~:
\begin{eqnarray}
\left\{ \begin{array}{c} A_{K^0 \Sigma^+} \\ 
B_{K^0 \Sigma^+} \\ \end{array} \right\} 
&&= \int_{-1}^1 dx \,
\left\{ \begin{array}{c} \widetilde H^{p \to \Sigma^+} \\ 
\widetilde E^{p \to \Sigma^+} \\ \end{array} \right\} \nonumber\\ 
\times&&\left[ {{(1 - \eta^a_K / \eta^s_K) \, e_d}
 \over {x - \xi + i \epsilon}} 
+  {{(1 + \eta^a_K / \eta^s_K) \,  e_s }
\over {x + \xi - i \epsilon}} \right] . 
\label{eq:abko}
\end{eqnarray}
In contrast to $\pi^0$ electroproduction, $K^0$ electroproduction 
can contain a pole contribution,
which is given by~: 
\begin{equation}
B^{pole}_{K^0 \Sigma^+} \,=\, {4 \over 3} \, \eta^a_K \,
\left( {3 \over {2 \xi}} \right) \, 
{{f_K \, g_{K N \Sigma} \, ({2 m_N})} \over {-t + m_K^2}} \;,
\label{eq:kosigpole}
\end{equation}
and which vanishes when the kaon DA is symmetric (i.e. when $\eta^a_K$ = 0).
Therefore, the $K^0$ pole contribution to $B_{K^0 \Sigma^+}$, provides
a direct measure of the antisymmetric component of the kaon DA.
\newline
\indent
In Fig.~\ref{fig:pionkhyp}, we compare the leading order 
predictions for pion and kaon hard electroproduction reactions 
at $Q^2$ = 10 GeV$^2$. 
The charged pion and kaon channels obtain a
large contribution in the range $x_B \gtrsim 0.1$ from the pion (kaon)
pole. This largely determines the ratio between these
channels at larger $x_B$. For values of $-t$ in the range 
0.1 $\to$ 0.5 GeV$^2$, this yields 
$\pi^+ n : K^+ \Lambda \approx 7 : 1 \to 1.8 : 1$, 
using an asymptotic DA for both $\pi$ and $K$. The kaon DA is
not well known however, and the results with a CZ kaon DA yield $K^+$ cross
sections larger by a factor (7/5)$^4 \approx$ 3.8, for the pole
contribution.    
The ratio  $K^+ \Lambda : K^+ \Sigma^0$ at large $x_B$ 
is determined from the ratio of
the couplings : $g^2_{K N \Lambda} / g^2_{K N \Sigma} \approx 12$.
For the $K^0 \Sigma^+$ channel, the pole contribution is absent if
$\eta^a_K$ = 0 (as for $\pi^0$). 
In this case, the ratio $\pi^0 p : K^0 \Sigma^+$
is determined by the PV contribution and is
very sensitive to the input valence quark distribution into $\tilde
H$. For $\Delta u_V \approx - \Delta d_V$ expected in the large $N_c$
limit, $\pi^0 : K^0 \approx 1 : 3$, while for 
$\Delta u_V \approx -2 \Delta d_V$ prefered by the global fit to DIS of 
Ref.~\cite{Leader},  $\pi^0 : K^0 \approx 3 : 1$. The
sensitivity of this ratio to the polarized quark distributions 
might be interesting to provide cross-checks on such
global fits from DIS. In Fig.~\ref{fig:pionkhyp}, we show the results for 
$K^0 \Sigma^+$ by using the polarized distributions of \cite{Leader} as
input for $\tilde H$ (as in \cite{VGG99}). 
Besides the PV contribution, $K^0 \Sigma^+$ 
electroproduction has also a pole contribution, if $\eta^a_K \neq 0$. 
We include the pole contribution of Eq.~(\ref{eq:kosigpole}), and use 
the CZ kaon DA with $\eta_K^a$ = 0.25. 
The resulting $K^0$ pole contribution
provides a sizeable enhancement of the $K^0 \Sigma^+$ cross section (it is
roughly half the value of the PV contribution at the largest $x_B$). 
\vspace{.3cm}
\begin{figure}[h]
\epsfxsize=7.5cm
\epsfysize=8.5cm
\centerline{\epsffile{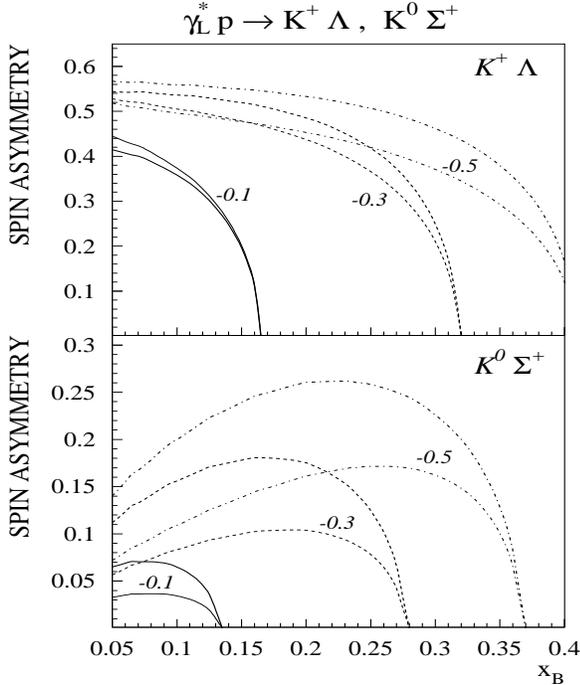}}
\caption[]{\small Transverse spin asymmetry, 
  for $K^+ \Lambda$ and $K^0 \Sigma^+$ longitudinal
  electroproduction for different values of $t$ (indicated on the
  curves in (GeV/c)$^2$). For $K^+ \Lambda$, 
thick (thin) lines are the predictions with 
asymptotic (CZ) kaon DA. For $K^0 \Sigma^+$,
  thick (thin)  lines are the predictions with CZ type kaon DA, with 
  antisymmetric part : $\eta^a_K$ = 0.25 (0.1).} 
\label{fig:khypasymm}
\end{figure}
\indent
For the strangeness channels, ${\cal A}_{K Y}$ is given by~: 
\begin{eqnarray}
&&{\cal A}_{K Y} = {{2 \, |\Delta_\perp|} \over {\pi}} \,
\frac{Im(A B^*) \, 4 \xi m_N}{D_{K Y}} \, , \\
&&D_{K Y} = |A|^2  4 m_N^2 (1-\xi^2) \,+\,
|B|^2 \xi^2 \left[ - t + (m_Y - m_N)^2 \right] \nonumber\\
&&- \, Re(AB^*) \, 4 \xi m_N
\left[ \xi (m_Y + m_N) + m_Y - m_N \right] \, , \nonumber
\end{eqnarray}
where $A$ and $B$ are as defined before. 
Interestingly, that in the case of hyperon production,
the same interference azimuthal spin asymmetry can be measured on an 
{\it unpolarized} target by measuring the polarization 
of the recoiling hyperon through its decay angular distribution.  
${\cal A}_{K^+\Lambda},{\cal A}_{K^0 \Sigma^+}$ 
are shown in Fig.~\ref{fig:khypasymm}. They are
 as large as 
for $\pi^+ n$. We also find 
${\cal A}_{K^+\Sigma^0}\sim {\cal A}_{K^+\Lambda}$.
For $K^0$ production, the sensitivity to the 
$SU(3)_f$ symmetry breaking effects in the kaon DA 
is illustrated (lower panel of Fig.~\ref{fig:khypasymm}),
by plotting ${\cal A}_{K^0 \Sigma^+}$ for two values of $\eta^a_K$. 
Because ${\cal A}_{K^0 \Sigma^+}$ is directly
proportional to $\eta^a_K$, it provides a very sensitive observable
to extract the $K^0$ form factor.
\newline
\indent
To summarize, we have shown that yields for hard exclusive 
production of decuplet and octet baryons are similar. 
Strange and nonstrange 
channels can be comparable and in some channels strange can even dominate
(depending on DA and polarized parton distributions),   
in contrast to low-energy strangeness production. 
Large transverse spin asymmetries are predicted for many of
these reactions. Several tests of validity of the large $N_c$ approximation in 
QCD would be possible.
\newline
\indent
This work was supported by the Israeli Academy of Science, 
a US DOE grant, the Alexander von Humboldt Foundation,  
the BMBF and the Deutsche Forschungsgemeinschaft (SFB443).

\end{document}